\documentclass[aps,amsfonts,amsmath,prb,twocolumn]{revtex4}

\usepackage{graphicx}  
\usepackage{array}
\begin{document}
\title{Size selection and stability of thick-walled vesicles}
\date\today
\author{M.~J.~Greenall}

\affiliation{School of Mathematics and Physics,
    University of Lincoln, Brayford Pool, Lincoln LN6 7TS}

\begin{abstract}
In recent experiments, small, thick-walled vesicles with a narrow size distribution were formed from copolymers where the degree of polymerisation of the hydrophobic
block, $N_\mathrm{B}$, was significantly greater than that of the hydrophilic block, $N_\mathrm{A}$. Using a mean-field theory, we reproduce several aspects of the behaviour of these vesicles. Firstly, we find a minimum in the free energy of the system of vesicles as a function of their radius, corresponding to a preferred size for the vesicles, when $N_\mathrm{B}$ is several times larger than
$N_\mathrm{A}$. Furthermore, the vesicle radius diverges as $N_\mathrm{B}$ is increased towards a critical value, consistent with the instability of the vesicles with respect to further aggregation seen in the experimental work. Finally, we find that this instability can also be triggered in our model by changing the interaction strength of the copolymers with the solvent.
\end{abstract}

\maketitle

\section{Introduction}
Amphiphiles such as lipids and block copolymers can
self-assemble into a range of aggregates in a solvent \cite{hamley_book}. Some of
these structures, known as {\it micelles}, consist of a hydrophobic core surrounded by a
hydrophilic corona. Micelles may be spherical or worm-like in
shape \cite{derry}. Other aggregates, called {\it
  vesicles}, are bag-like structures formed of a {\it bilayer} of
amphiphiles, and enclose a volume of solvent. Although
vesicles are often roughly spherical \cite{warren}, their
underlying physics differs from that of
spherical micelles. In particular, the radius of a spherical micelle
can be predicted from the architecture and interactions of the amphiphiles \cite{lund,LOW}. However, the size of a vesicle is often
determined by other factors. In
systems containing only one type of simple amphiphile (such as a lipid
or diblock copolymer), the effect that limits
the growth of self-assembled vesicles is often their translational
entropy \cite{simons_cates,coldren}. The resulting
vesicle size distribution is often broad \cite{enders, li} and depends
sensitively on the amphiphile concentration
\cite{morse_milner}. In practice, control over the size of vesicles is
often obtained by filtration \cite{storslett} or by using a more
complex preparation method, such as dewetting from a template
\cite{howse}. The possibility also exists of mixing two types of
amphiphile \cite{andelman}, which divide unevenly between the inner
and outer leaflets of the membrane and give the vesicle a preferred size.

However, in recent experiments \cite{warren}, vesicles with a
narrow size distribution have been formed in a solution of a
single type of diblock copolymer by self-assembly. Vesicles with comparable
size distributions were formed by two different
pathways: polymerisation-induced self-assembly (PISA), where
polymerisation continues after self-assembly has started, and
rehydration of a thin copolymer film \cite{warren}. This suggests that
a suspension of vesicles of a well-defined size is the equilibrium
phase in this system. These vesicles have two distinguishing
features. Firstly, they are formed of highly asymmetric polymers, with
the degree of polymerisation of the hydrophobic block as high as
15 times that of the hydrophilic block \cite{warren}. Secondly, the walls
of the vesicles are thick, and their thickness is often of the order
of magnitude of the radius of the central liquid pocket. In this report, we present a mean-field model that predicts the the existence of vesicles with a preferred
size for strongly asymmetric copolymers and also reproduces other
aspects of the experimental system.

\section{Mean-field model}
Mean-field models have been used to study
spherical micelles in solution
\cite{LOW,noolandi_hong,whitmore_noolandi}, and have also been applied
to cylindrical micelles \cite{mayes_delacruz} and flat bilayers
\cite{munch_gast}. They provide a good description of experimental
results on both block copolymers in solution \cite{lund} and block
copolymer/homopolymer blends \cite{roe}. To set up a mean-field model,
the principal contributions to the free energy of a system of micelles
(neglecting fluctuations)
are identified, and formulas for these are found. For example, the
Flory-Huggins expression is used for the free energy of mixing of
copolymers and homopolymers outside the micelles \cite{LOW}. The
various terms are then added together, and the resulting expression is
minimised. This
yields a number of predictions,
including the equilibrium radius of the micelles. Here, we apply this approach to a system of spherical
vesicles, with the aim of finding whether these aggregates have a
preferred radius at equilibrium. The theory will be developed for diblock
copolymers formed of $N_\mathrm{A}$ A monomers and
$N_\mathrm{B}$ B monomers mixed with a `solvent' of
homopolymers containing $N_\mathrm{h}$ A monomers. However, for
simplicity, we will set $N_\mathrm{h}=1$ in our numerical calculations.
\begin{figure}[ht]
\begin{center}
\includegraphics[width=0.4\textwidth]{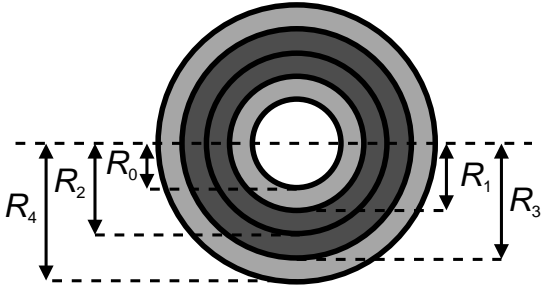}
\caption{\label{cross_section} Geometry of a spherical bilayer vesicle with
  outer radius $R_4$. The
  dark grey regions are hydrophobic and the light grey regions are
  hydrophilic. The vesicle is surrounded by solvent and encloses a
  spherical volume of solvent with radius $R_0$ in its centre.}
\end{center}
\end{figure}
To begin, we introduce the
contributions to the free energy of a single vesicle. The first of
these arises from the fact that the copolymers in an aggregate are
deformed away from their unperturbed state \cite{LOW}. This leads to
an elastic energy term for the inner leaflet of the vesicle given by
\begin{multline}
\mathcal{F}_\mathrm{d}^\mathrm{in}=\frac{3}{2}kTp_1\left\{
\frac{(R_1-R_0)^2}{N_\mathrm{A}a^2}
+\frac{N_\mathrm{A}a^2}{(R_1-R_0)^2}\right.\\
\left.+\frac{(R_2-R_1)^2}{N_\mathrm{B}a^2}
+\frac{N_\mathrm{B}a^2}{(R_2-R_1)^2}
-4\right\},
\label{deformation}
\end{multline}
where $k$ is Boltzmann's constant, $T$ is the temperature, $a$ is the
segment length, $p_1$ is
the number of copolymers in the inner leaflet and the $R_i$ are
the radii shown in Fig.\ \ref{cross_section}. This term is zero when the polymers are in their unperturbed state and gives an
energy penalty when they are stretched or compressed
\cite{LOW, roe,degennes}. A similar term,
$\mathcal{F}_\mathrm{d}^\mathrm{out}$, exists for the outer leaflet,
which contains $p_2$ copolymer chains.

The hydrophilic layers of the vesicle (light grey in Fig.\ \ref{cross_section}) are assumed to consist of copolymer A blocks and solvent, and the hydrophobic layers
(dark grey in Fig.\ \ref{cross_section})
of copolymer B blocks and solvent. This leads to a term
representing the entropy of mixing of the solvent with the copolymer
chains \cite{LOW}, given in the case of a vesicle by
\begin{equation}
\mathcal{F}_\mathrm{m}=\sum_{i=1}^4\frac{4\pi}{3} \frac{R^3_i-R^3_{i-1}}{a^3}kT\frac{1-\eta_i}{N_\mathrm{h}}\ln(1-\eta_i).
\end{equation}
Here, the volume fraction of
copolymer in each layer is $\eta_i$, where $i$
runs from 1 in the
innermost layer to 4 in the outermost layer.

The solvent has a repulsive interaction with the hydrophobic blocks, whose strength is given by the Flory-Huggins $\chi$
parameter. This results \cite{lund} in the following term in the free energy:
\begin{multline}
\mathcal{F}_\mathrm{core}=\frac{4\pi}{3}\frac{(R^3_2-R_1^3)}{a^3}kT\eta_2(1-\eta_2)\chi\\
+\frac{4\pi}{3}\frac{(R^3_3-R^3_2)}{a^3}kT\eta_3(1-\eta_3)\chi.
\end{multline}
The vesicle contains two surfaces that separate a hydrophilic region
from a predominantly hydrophobic region. Each of these produces a
contribution to the free energy of the vesicle proportional to
its area and to the square root of the $\chi$ parameter \cite{LOW,helfand_tagami}:
\begin{equation}
\mathcal{F}_\mathrm{int}=4\pi
  R^2_1\frac{kT}{a^2}\sqrt{\frac{\chi}{6}}\eta_2+4\pi R^2_3\frac{kT}{a^2}\sqrt{\frac{\chi}{6}}\eta_3.
\end{equation}
The factors of $\eta_2$ and $\eta_3$ arise since the hydrophobic
layers contain some solvent, and each term is reduced from the value
it would have for an interface between pure hydrophobic and
hydrophilic layers \cite{lund}. The total free energy of a vesicle is
given by the sum of the terms above:
$\mathcal{F}=\mathcal{F}_\mathrm{d}^\mathrm{in}+\mathcal{F}_\mathrm{d}^\mathrm{out}+\mathcal{F}_\mathrm{m}+\mathcal{F}_\mathrm{int}$. We
also assume that the system is incompressible. This allows us to
express the $\eta_i$ in terms of the copolymer parameters and the
dimensions of the vesicle and so reduce the number of variables. For
example, in the innermost layer, $\eta_1=3p_1N_\mathrm{A}a^3/[4\pi
(R^3_1-R^3_0)]$.

To calculate the free energy of a system of vesicles, we note that, if $\Omega$ is the total number of monomers in
the system, $\phi$ is the volume fraction of copolymers and $\zeta$ is
the fraction of copolymer chains in aggregates, then the total number of vesicles is given by
$\Omega\phi\zeta/[(p_1+p_2)(N_\mathrm{A}+N_\mathrm{B})]$. We can then
write the total free energy of the system as
\begin{equation}
F_\mathrm{M}=\{\Omega\phi\zeta/[(p_1+p_2)(N_\mathrm{A}+N_\mathrm{B})]\}\mathcal{F}+F_\mathrm{mix}-TS_\mathrm{m},
\label{total}
\end{equation}
where $F_\mathrm{mix}$ is the free energy of mixing of copolymers and
solvent outside the vesicles \cite{roe_zin} and $S_\mathrm{m}$ is the translational
entropy of the `gas' of vesicles \cite{LOW}. Adapting the expression
in Ref.\ \citenum{LOW} to the case of vesicles, we find that the free energy of mixing 
is given by
\begin{multline}
\frac{F_\mathrm{mix}}{kT}=\Omega(1-\xi\phi\zeta)\left[
\frac{\phi_1}{N}\ln\phi_1+
\frac{1-\phi_1}{N_\mathrm{h}}\ln(1-\phi_1)\right.\\
\left.+\frac{\chi N_\mathrm{B}\phi_1}{N_\mathrm{A}+N_\mathrm{B}}\left(1-\frac{\phi_1
  N_\mathrm{B}}{N_\mathrm{A}+N_\mathrm{B}}\right)
\right],
\label{mix}
\end{multline}
where
\begin{multline*}
\xi=\frac{1}{(p_1+p_2)(N_\mathrm{A}+N_\mathrm{B})}\left(\frac{p_1N_\mathrm{A}}{\eta_1}+\frac{p_1N_\mathrm{B}}{\eta_2}\right.\\
\left.+\frac{p_2N_\mathrm{B}}{\eta_3}+\frac{p_2N_\mathrm{A}}{\eta_4}\right).
\end{multline*}
The factor of $\Omega(1-\xi\phi\zeta)$ in Eqn.\ \ref{mix} is the total
number of monomers outside the vesicles, and
$\phi_1=\phi(1-\zeta)/(1-\xi\phi\zeta)$ is the fraction of monomers
outside vesicles that belong to copolymers. Similarly, we adapt the lattice
model calculation of the translational entropy of micelles in
Ref. \citenum{LOW} to the case of vesicles, and find that 
\begin{multline}
\frac{S_\mathrm{m}}{k}=-\Omega\left\{\frac{\phi\zeta}{(p_1+p_2)(N_\mathrm{A}+N_\mathrm{B})}\ln(\phi\zeta\tilde{\xi})\right.\\
\left.+\frac{1-\phi\zeta\tilde{\xi}}{\tilde{\xi}(p_1+p_2)(N_\mathrm{A}+N_\mathrm{B})}\ln(1-\phi\zeta\tilde{\xi})\right\},
\end{multline}
where
\begin{equation*}
\tilde{\xi}=4\pi R_4^3/[(p_1+p_2)(N_\mathrm{A}+N_\mathrm{B})3a^3].
\end{equation*}
The differences with the micelle calculation arise from the existence of the two layers in the
vesicle wall and the fact that the central pocket of solvent must be
treated as being within the vesicle.

To find whether the vesicles formed from polymers with a given set of
values for $N_\mathrm{A}$, $N_\mathrm{B}$ and $\chi$ have a preferred
size, we first set $R_4$ to a fixed value and minimise Eqn.\
\ref{total} with respect to $R_0$, $R_1$, $R_2$, $R_3$, $p_1$, $p_2$
and $\phi_1$ using a direction set method \cite{num_rec}. We then
repeat the calculation for different values of $R_4$ to find whether
$F_\mathrm{M}$ has a minimum as a function of $R_4$ (corresponding to a
preferred radius).

\section{Results}
\begin{figure}[ht]
\begin{center}
\includegraphics[width=0.4\textwidth]{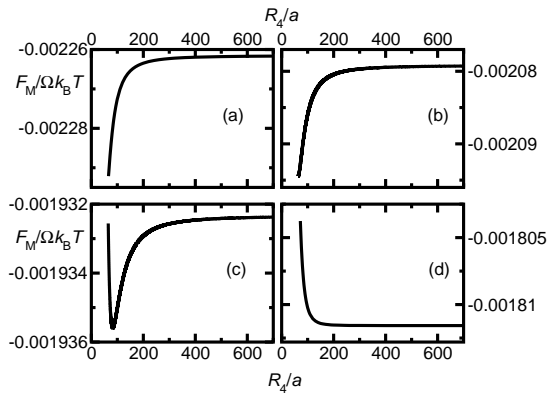}
\caption{\label{first_min} Total free energy of the system of
  vesicles versus the outer radius of the vesicle at $N_\mathrm{A}=100$ and (a)  $N_\mathrm{B}=300$; (b)
  $N_\mathrm{B}=350$; (c) $N_\mathrm{B}=400$; (d) $N_\mathrm{B}=450$.}
\end{center}
\end{figure}

To begin, we focus on a system of relatively short copolymers, and
set $N_\mathrm{A}=100$ while varying $N_\mathrm{B}$. Since the solvent
consists of A monomers (so that $N_\mathrm{h}=1$), we set $\chi$ to the
relatively high value of $2$ to ensure that aggregation takes place
over a range of $N_\mathrm{B}$. The volume fraction of
copolymers is set to $\phi=0.01$, giving a dilute system. Plots of
$F_\mathrm{M}/\Omega k_\mathrm{B}T$ are shown in Fig.\
\ref{first_min}. For $N_\mathrm{B}\lesssim 350$, $F_\mathrm{M}$ falls
monotonically as $R_4$ decreases, dropping sharply for $R_4\lesssim
200$. At small values of $R_4$, we are no longer able to minimise Eqn.\
\ref{total}. A likely explanation of these results is
that the vesicle is unstable with respect to micelle
formation. This is borne out by the fact that $p_1$ shrinks rapidly as $R_4$ becomes small.
In contrast, when $N_\mathrm{B}=400$, a clear minimum is
present in the free energy, corresponding to a preferred size for the
vesicles. When $N_\mathrm{B}$ is increased to $450$, the minimum
disappears, and $F_\mathrm{M}$ decays monotonically as $R_4$
increases. Here, there is no optimum radius,
and the system might either precipitate or form vesicles with a broad size
distribution.

\begin{figure}[ht]
\begin{center}
\includegraphics[width=0.4\textwidth]{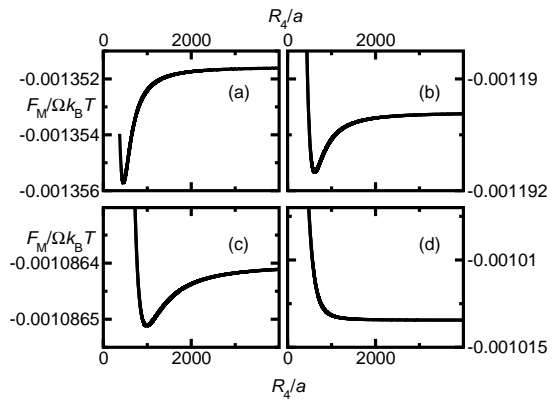}
\caption{\label{second_min} The total free energy of the system of
  vesicles as a function of the outer radius of the vesicle at fixed hydrophilic block length $N_\mathrm{A}=1000$ and four
  different hydrophobic block lengths: (a)  $N_\mathrm{B}=10000$; (b)
  $N_\mathrm{B}=13000$; (c) $N_\mathrm{B}=16000$; (d) $N_\mathrm{B}=19000$.}
\end{center}
\end{figure}

Next, we consider longer copolymers, with
$N_\mathrm{A}=1000$ and $N_\mathrm{B}$ being varied (Fig.\ \ref{second_min}). Here, the minimum first appears for more strongly
asymmetric copolymers, with $N_\mathrm{B}\approx 10N_\mathrm{A}$, and
persists over a wider range of values of $N_\mathrm{B}$. At all free
energy minima
shown here (and in Fig. \ref{first_min}c above), the vesicles are
thick-walled, with a small central solvent pocket. For example, in Fig.\ \ref{second_min}c, when $N_\mathrm{B}=16000$ the outer radius of the vesicle
at the minimum is $R_4\approx 989a$, while the inner radius is
$R_0\approx 382a$. Having found a favoured vesicle size in
calculations on two families of copolymers, we now attempt to understand the
physical processes that lead to this effect.
\begin{figure}[ht]
\begin{center}
\includegraphics[width=0.4\textwidth]{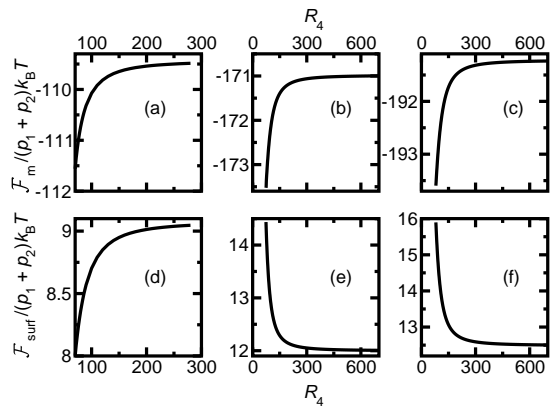}
\caption{\label{mix_vs_surf} Top three panels: free energy of
mixing per chain in the vesicle versus the vesicle outer radius for
(a) $N_\mathrm{A}=100,\, N_\mathrm{B}=100$; (b) $N_\mathrm{A}=100,\,
N_\mathrm{B}=400$; (c) $N_\mathrm{A}=100,\, N_\mathrm{B}=500$. Bottom
three panels: surface free energy of the vesicle per
chain versus the vesicle outer radius for (d) $N_\mathrm{A}=100,\, N_\mathrm{B}=100$; (e) $N_\mathrm{A}=100,\,
N_\mathrm{B}=400$; (f) $N_\mathrm{A}=100,\, N_\mathrm{B}=500$. }
\end{center}
\end{figure}

As $R_4$ is varied, the contributions to the free energy that vary
over the largest range are those associated with the
individual vesicles,
with the bulk contributions changing
more slowly. This means that, as for micelles \cite{LOW},
the equilibrium structure of the vesicle may be found, to a good
approximation, by minimising the free energy per chain in the
vesicle. Of the terms in the free energy per chain in the vesicle,
those that vary most are the free energy of mixing, $\mathcal{F}_\mathrm{m}$, and
the surface free energy, $\mathcal{F}_\mathrm{surf}$, and these are
plotted in Fig.\ \ref{mix_vs_surf} for $N_\mathrm{A}=100$ and a range
of values of $N_\mathrm{B}$. We see that the free energy of mixing per
chain rises as $R_4$ is increased for all the values of $N_\mathrm{B}$
shown. When $N_\mathrm{B}=100$, the surface free energy per chain also
grows as $R_4$ is increased. However, for $N_\mathrm{B}=400$, it
becomes a decreasing function of $R_4$. The fall in the surface
free energy per chain is very close to the rise in the free energy of
mixing per chain,
and this fine balance leads to the minimum in
$F_\mathrm{M}$ in Fig.\ \ref{first_min}c. In the final
case, when $N_\mathrm{B}=500$, the fall in the surface energy is faster, and dominates the rise in the mixing energy,
leading to a monotonic decay of $F_\mathrm{M}$ with $R_4$.
\begin{figure}[ht]
\begin{center}
\includegraphics[width=0.4\textwidth]{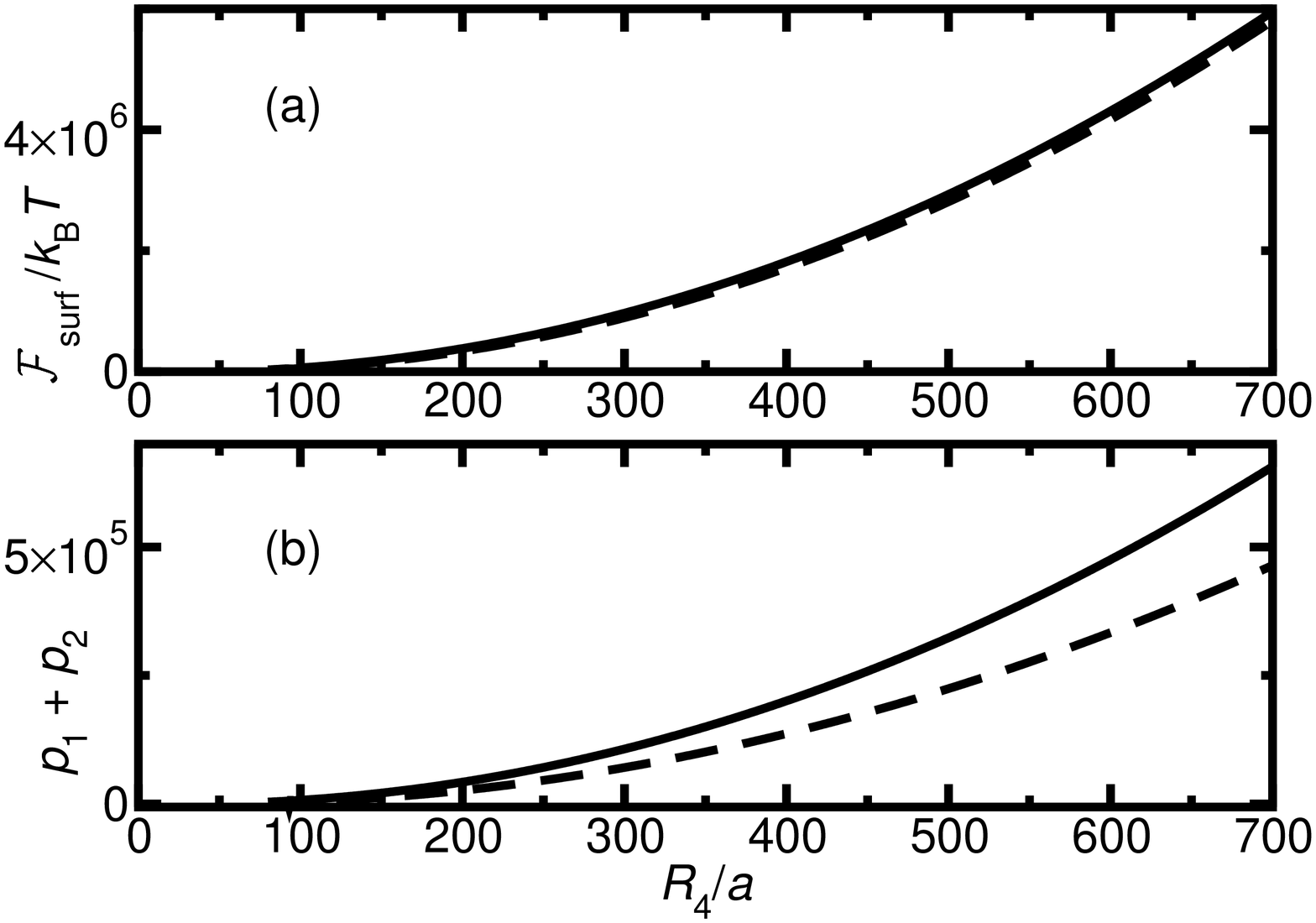}
\caption{\label{surfp1p2} (a) Vesicle surface free energy versus outer radius for
  $N_\mathrm{A}=100,\, N_\mathrm{B}=100$ (full line) and
  $N_\mathrm{A}=100,\, N_\mathrm{B}=500$ (dashed line). (b)
  Vesicle aggregation number versus outer radius for $N_\mathrm{A}=100,\, N_\mathrm{B}=100$ (full line) and
  $N_\mathrm{A}=100,\, N_\mathrm{B}=500$ (dashed line).}
\end{center}
\end{figure}
Having identified the change in behaviour of the surface free energy
per chain as the mechanism behind the appearance and disappearance of
size selection as $N_\mathrm{B}$ is increased, we now study
this term in more detail and plot the surface free energy and the
aggregation number of the vesicle separately in Fig.\
\ref{surfp1p2} for copolymers with $N_\mathrm{A}=100,\, N_\mathrm{B}=100$
and $N_\mathrm{A}=100,\, N_\mathrm{B}=500$. As would be expected, $\mathcal{F}_\mathrm{surf}$ 
for the longer copolymers is slightly lower at a given $R_4$, as the vesicle walls are thicker and the radius of the inner
hydrophobic/hydrophilic interface, $R_1$, is smaller. Both curves can be fitted very well by a power law of the form
$F_\mathrm{surf}/k_\mathrm{B}T\sim R_4^\rho$ with $\rho\approx 2.139$
when $N_\mathrm{B}=100$ and $\rho\approx 2.200$ when
$N_\mathrm{B}=500$. 

The difference between the aggregation numbers of the two vesicles is more
pronounced, with that of the $N_\mathrm{B}=500$
vesicles being smaller due to the greater volume of the longer molecules. Again, both sets of results can be
fitted by a power law, so that $p_1+p_2\sim R_4^\sigma$ with $\sigma=2.137$ when 
$N_\mathrm{B}=100$ and $\sigma=2.207$ when
$N_\mathrm{B}=500$. The stronger variation when $N_\mathrm{B}=500$ occurs since the bilayers of these vesicles are highly
asymmetric at small $R_4$, with $p_2\gg p_1$. As $R_4$ increases, more molecules enter the inner
leaflet, and the asymmetry decreases. In constrast, the
$N_\mathrm{B}=100$ vesicles are close to
symmetric at smaller values of $R_4$ and so display a weaker variation
of $p_1+p_2$ with $R_4$.

Since $\rho>\sigma$ when $N_\mathrm{B}=100$, but $\sigma>\rho$ when $N_\mathrm{B}=500$, the surface free energy per chain changes from an
increasing function of $R_4$ to a decreasing function as
$N_\mathrm{B}$ is increased. The fine balance between these terms
leads to the appearance of vesicles of a preferred size for a range of
$N_\mathrm{B}>N_\mathrm{A}$, which ultimately become unstable for
large $N_\mathrm{B}$.

\begin{figure}[ht]
\begin{center}
\includegraphics[width=0.4\textwidth]{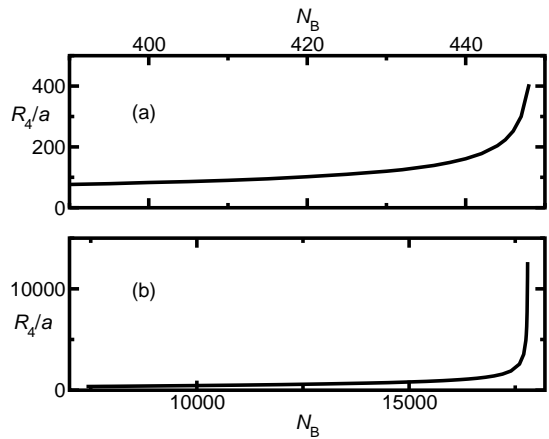}
\caption{\label{radii} Outer radius of the vesicle versus
  the degree of polymerisation of the hydrophobic block for (a)
  $N_\mathrm{A}=100$ and (b) $N_\mathrm{A}=1000$.}
\end{center}
\end{figure}

We now look in more detail at the growth of the vesicle as
$N_\mathrm{B}$ is increased. In Fig.\ \ref{radii}, we plot $R_4$ against $N_\mathrm{B}$ for
(a) $N_\mathrm{A}=100$ and (b) $N_\mathrm{A}=1000$. In both cases, the vesicle
radius grows slowly at first before diverging as a critical value of
$N_\mathrm{B}$. This is consistent with PISA
experiments \cite{derry,warren}, where the vesicles become unstable
above a critical value of $N_\mathrm{B}$.

\begin{figure}[ht]
\begin{center}
\includegraphics[width=0.4\textwidth]{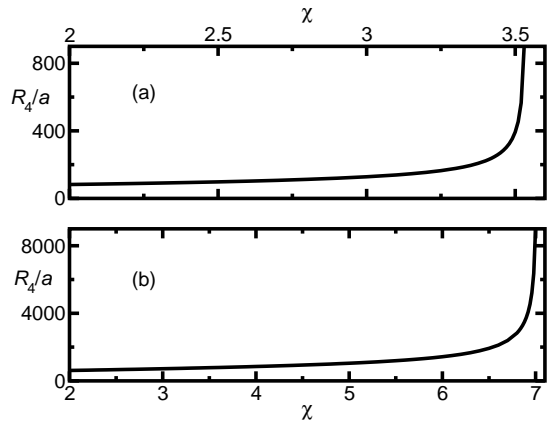}
\caption{\label{radii_chi} Outer
  radius of the vesicle versus $\chi$ parameter for (a) $N_\mathrm{A}=100,\,
  N_\mathrm{B}=400$ and (b) $N_\mathrm{A}=1000,\, N_\mathrm{B}=13000$.}
\end{center}
\end{figure}

The instability can also be triggered by a change in $\chi$. In Fig.\
\ref{radii_chi}, we plot $R_4$ against $\chi$ for the (a)
$N_\mathrm{A}=100,\, N_\mathrm{B}=400$ and (b) $N_\mathrm{A}=1000,\,
N_\mathrm{B}=13000$ copolymers. The radius is initially relatively
insensitive to $\chi$ for both copolymers, before diverging sharply at
$\chi\approx 3.5$ in the case of the shorter molecules and
$\chi\approx 7$ for the longer ones.

\section{Conclusions}
Using a mean-field model, we have reproduced a number of features of
the small, thick-walled vesicles formed in recent experiments \cite{derry,warren}. Our model predicts that the vesicles have
a preferred radius for a range of parameters where the
hydrophobic block is much longer than the hydrophilic block, agreeing with the experimental observation of a
narrow vesicle size distribution in solutions of such polymers
\cite{warren}. In our calculations, the origin of the free energy
minimum that leads to the existence of a preferred radius is found to
be a competition between the free energy of mixing of the solvent with
the copolymers in the vesicle and the surface free energy of the vesicle. For asymmetric polymers with a long hydrophobic block, these terms are
finely balanced, and a minimum in the free energy as a function of the
vesicle radius appears. If the hydrophobic block is shortened, the
minimum disappears, and the vesicles become unstable with respect to
the formation of smaller structures; i.e., spherical micelles. If, on
the other hand, the
hydrophobic block is lengthened, the free energy becomes a
monotonically decreasing function of the vesicle radius, so that the
vesicles no longer have a preferred size and may be, as in the experiments, unstable with
respect to further aggregation. We also find that this instability may
be triggered by a change in the interaction of the vesicles with the
surrounding solvent. This raises the possibility of vesicles that
burst, perhaps releasing an encapsulated cargo, when they move into a
particular chemical environment.

Several extensions to our work are possible. Firstly, more
realistic model parameters, and/or the use of different $\chi$ parameters
for the A-block/B-block, A-block/solvent and B-block/solvent
interactions, might improve the agreement of
our theory with experimental results. In particular, our model
predicts an initial slow growth of the vesicle radius with
$N_\mathrm{B}$ that accelerates gradually as the instability is
approached. This is not in complete agreement with the experimental results
\cite{derry}, where the radius of the aggregates remains essentially constant over a range of
$N_\mathrm{B}$ before increasing sharply at the instability. Finally, interdigitation of the molecules could be incorporated in a
modified theory to allow the modelling of a wider range of
systems.
\bibliography{Mean_fieldrefs} 

\end{document}